# Experimental Characterization of Non-Associative Plasticity Flow Rule Coefficients for the LS-DYNA MAT213 Model

Ryan Premo[1] and Jackob Black[2]
*Mississippi State University, Mississippi State, MS 39762, USA*

Michael Pereira[3], Robert K. Goldberg[4], and Trenton M. Ricks[5]
*NASA Glenn Research Center, Cleveland, OH 44135, USA*

Han-Gyu Kim[6]
*Mississippi State University, Mississippi State, MS 39762, USA*

**This project is focused on developing an experimental framework for characterizing non-associative plasticity flow rule coefficients through coupon-scale tests for the LS-DYNA MAT213 model. The main objective is to characterize these coefficients based on the multi-scale (i.e., both microscopic and macroscopic) full-field measurement of the evolution of strain and stress fields. This paper focuses on presenting the experimental work on characterizing the full-scale stress-strain curves of T700/LM-PAEK composites under tension, compression, and shear loads. The experimental data set was intended to build a deformation sub-model in the MAT213 model for the material. The strain data were collected using both microscopic and macroscopic digital image correlation techniques. The microscopic technique was particularly useful for fracture cases under small strains. A preliminary simulation result obtained from the MAT213 model is also presented in the paper. The experimental framework herein will be extended to characterize post-peak stress degradation in the composite material and to develop a damage sub-model for the material. This project will contribute to developing a simulation tool based on the MAT213 model for simulating the rate-dependent impact damage in composites under multi-axial loading.**

## I. Introduction

LS-DYNA is a commercially available transient dynamic finite element solver and utilizes various material models depending on the application and simulation desired. For implementation into LS-DYNA, the MAT213 material model has been developed by a consortium led by the Federal Aviation Administration (FAA) and the National Aeronautics and Space Administration (NASA). The MAT213 model simulates deformation and damage progression in composite structures subjected to high-speed impacts [1]. The deformation sub-model in the MAT213 model requires several material parameters as inputs to define the non-associative plastic flow rule of the orthotropic plasticity model. In the current methodology, however, complex material properties such as plastic Poisson's ratios need to be measured to determine some of the key parameters [2]. Furthermore, for simulations of post-peak stress degradation using the damage sub-model in the MAT213 model, the methodology requires correlation based on the structural-level impact or crush tests [3], which are extremely challenging with respect to cost and resources. To begin to address these issues, the current work is focused on developing an experimental framework for characterizing the non-associative plasticity flow rule coefficients through coupon-scale tests. The created framework will later be applied to the characterization of post-peak stress degradation, specifically for mode-II interlaminar fracture in composites under three-point bending. The main objective of the overall effort is to characterize these plasticity coefficients and stress degradation behaviors based on the multi-scale (i.e., both microscopic and macroscopic) full-field measurement of the evolution of strain and stress fields along fracture process zones in geometrically scaled specimens. The proposed methods are expected to enhance the fidelity of material parameter inputs for the MAT213 model while reducing the cost and resources required to obtain the parameters.

---

[1] Graduate Research Assistant, Department of Aerospace Engineering.
[2] Graduate Research Assistant, Department of Aerospace Engineering.
[3] Technical Lead, Ballistic Impact Lab.
[4] Research Aerospace Engineer, Ceramic and Polymer Composites Branch, Associate Fellow AIAA.
[5] Research Aerospace Engineer, Multiscale and Multiphysics Modeling Branch, Senior Member AIAA.
[6] Assistant Professor, Department of Aerospace Engineering. Corresponding author (hkim@ae.msstate.edu).

This paper primarily discusses the procedures required to conduct the required in-plane experimental tests, and the results obtained for the experiments. The modeling work on the acquired test data using the MAT213 model is still in progress; however, the preliminary results from single-element and coupon-level simulations in LS-DYNA utilizing the MAT213 model are briefly described in this paper. The chosen material for testing and modeling with LS-DYNA was the Toray Cetex® TC1225 low-melt polyaryl ether ketone (LM-PAEK) thermoplastic resin system, paired with T700G unidirectional carbon fibers, hereafter referred to as T700G/LM-PAEK. The National Center for Advanced Materials Performance (NCAMP) published a thorough lab testing report for the material properties of T700G/LM-PAEK [4]. The MAT213 model, however, requires tabulated, full-scale stress-strain curves as input, which necessitated the experimental work in this paper.

## II. Experimental Work

### A. Specimen details

As previously mentioned, specimens were manufactured using unidirectional T700G/LM-PAEK. One of the T700G/LM-PAEK panels is presented in Fig. 1. Before coupon cutting, C-scan images of the panels were captured for quality assurance as shown in Fig. 1b. The small white area to the left of the image indicates limited debonding induced by the manufacturing method. This small portion was marked on the panel (see Fig. 1a) and was discarded during coupon preparation. The test matrix for the specimens is described in Table 1. The tension, compression, and shear test sets in the table were intended to obtain the tabulated stress-strain curves required for input into MAT213. Additionally, the material properties listed in the table were used to validate the experimental data through comparison with the NCAMP data. The dimensions and geometries of the specimens for the test sets were designed based on the specifications published by ASTM International [5-8]. For the tension tests, tabs were attached to the specimens for robust clamping and were machined with a bevel angle of 7° to mitigate possible stress concentrations at the tab edges. The test fixture for microscopic tests did not allow for the additional thickness induced by tabbing. To address this issue, specimens were specially machined using a high-speed carbide bit in concert with an engineering mill to incorporate shims at the ends as shown in Fig. 2. Given the testing system size and the field of view (FOV) of the microscope, the gauge lengths and widths of these specimens were reduced from the ASTM specification maintaining the thickness values.

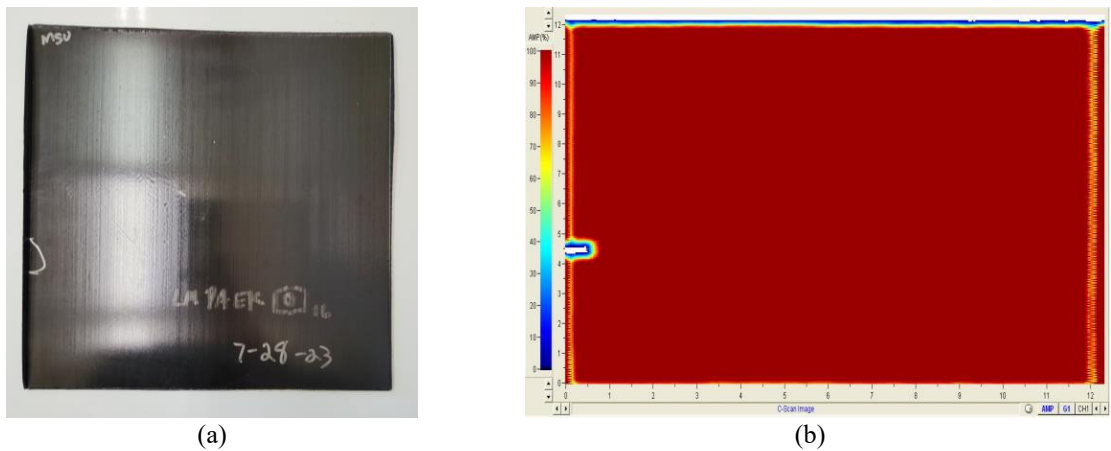

(a)            (b)

**Figure 1. A manufactured T700G/LM-PAEK panel. (a) An image of a T700G/LM-PAEK panel immediately prior to coupon cutting. (b) A C-scan image of the panel.**

Table 1. Test matrix

| Test type | Property type | Layup | ASTM specification |
|---|---|---|---|
| 0° unidirectional tension tests | 0° tensile modulus and strength | $[0]_8$ | D3039/D3039M-17 [5] |
| 90° unidirectional tension tests | 90° tensile modulus and strength | $[90]_{16}$ | D3039/D3039M-17 [5] |
| 0° unidirectional compression tests | 0° compressive modulus and strength | $[0]_{32}$ | D6641/D6641M-16e2 [6] |
| 90° unidirectional compression tests | 90° compressive modulus and strength | $[90]_{32}$ | D6641/D6641M-16e2 [6] |
| V-notched beam tests in 0° direction | 1-2 shear modulus and strength | $[0]_{16}$ | D5379/D5379M-19e1 [7] |
| V-notched beam tests in 90° direction | 2-1 shear modulus and strength | $[0]_{16}$ | D5379/D5379M-19e1 [7] |
| Tension tests of ±45° laminates | In-plane shear modulus | $[45/-45]_{2S}$ | D3518/D3518M-18 [8] |

**B. Experimental setup**

For multi-scale experimentation, two types of digital image correlation (DIC) systems were employed as shown in Fig. 3. A macroscopic DIC system (see Fig. 3a) provided coupon-scale data by capturing the stress-strain development on the entire specimen area. A Shimadzu AGX-V load frame and a Correlated Solutions VIC 2D package were employed for the coupon-scale tests. A microscopic DIC system (see Fig. 3b), on the other hand, was used to characterize fracture processes occurring at small strains in specimens such as the 90° unidirectional tension tests. For the microscopic tests, a Psylotech µTS testing frame, an Olympus BXFM microscope system with a 12-MP machine vision camera, and the VIC 2D package were employed. The FOV of the microscopic testing was 10 mm x 7.3 mm, and the lens was focused on the middle of the gauge section.

For the V-notched beam tests, the microscopic system was initially applied to obtain high-resolution data. The FOV available with the microscope, however, was not large enough to capture both V notches on the beams and shear failure could unpredictably be initiated from one of the notches. To address this issue, a hybrid experimental setup was built by merging the aspect of the macroscopic testing system with the µTS testing frame. In this testing configuration, the machine vision camera was paired with the macroscopic camera lens instead of the microscope and was cantilevered over the shear test fixtures. This combination allowed the complete capture of the coupon surface as shown in Fig. 4. An image of a speckled V-notched coupon in the test fixture shown in Fig. 4a was captured immediately preceding the shear fracture in the 90° direction. The DIC analysis of the image is presented for shear strains in Fig. 4b. The purple contours indicate large strains propagated from the notches in the 90° direction.

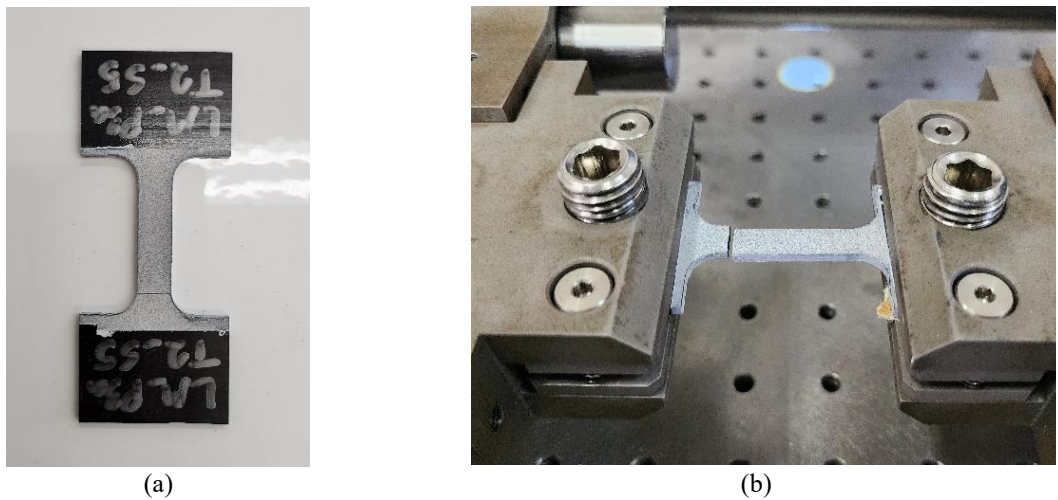

(a)　　　　　　　　　　　　　　　　　　(b)

**Figure 2. Microscopic coupon for the 90° unidirectional tension tests. (a) A dog-bone-shaped coupon incorporating shims at the ends. Note that this coupon has already been tested as such there is a fracture in the bottom portion of the gauge section. (b) The same coupon after fracture during a tensile test.**

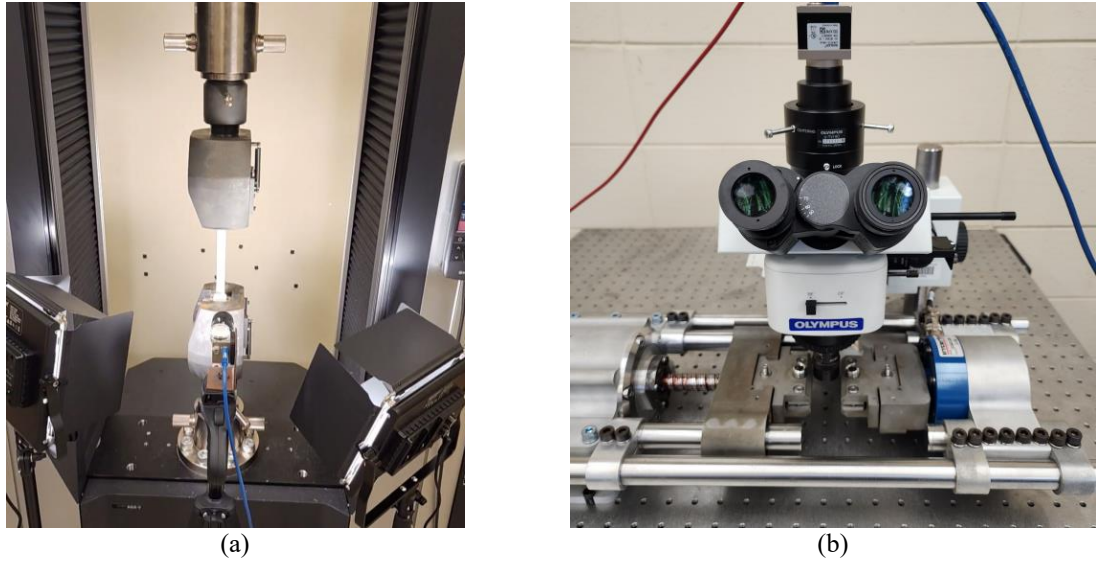

**Figure 3. Experimental setup for multi-scale DIC tests. (a) A Shimadzu AGX-V load frame for macroscopic DIC tests. (b) A Psylotech μTS testing frame for microscopic DIC tests.**

## III. Experimental Results

This experimental program was focused on acquiring full-scale stress-strain curves and material properties that could be utilized as inputs for the MAT213 material model. The material properties were also used to validate the experimental data using data obtained from the NCAMP report [4]. Each test set was considered to be complete when between three to five coupons had been tested, and the material's respective modulus and material strength both matched within the ±10% range of the NCAMP data. As shown in the test matrix (see Table 1), the complete set of in-plane macroscopic- and microscopic-scale tests encompassed unidirectional tension and compression tests in the 0° and 90° directions, V-notched beam tests in the 0° and 90° directions, and tension tests of ±45° laminates.

The full-scale stress-stress curves for longitudinal and shear strains are presented in Figs. 5 and 6. In this paper, the subscripts $xx$, $yy$, and $xy$ represent the longitudinal (i.e., the loading direction), transverse, and shear properties, respectively. Some of the stress-strain curves for the 0° unidirectional and ±45° tension tests (see Figs. 5a and 6c, respectively) showed sudden load drops due to the premature failures of singular fibers. This behavior was reduced and mitigated through subsequent improvements in the approach to coupon manufacturing and preparation (e.g., smoothing cutting surfaces, water cooling during panel cutting, etc.). Additionally, one of the 0° unidirectional tension specimens (see Fig. 5a) showed a smaller slope (i.e., the longitudinal modulus) than the other two, while one of the 0° unidirectional compression curves (see Fig. 5c) exhibited a different initial behavior from the other two but eventually exhibited a similar response. However, the other curves showed relatively consistent responses, and overall smooth stress-strain curves were obtained from all the test sets. This is particularly beneficial for the curve-fitting process of the MAT213 material model. Finally, for the 90° unidirectional tension tests, the stress-strain curves obtained from the microscopic and macroscopic data (see Fig. 5b) showed good agreement for longitudinal strains. As shown in Fig. 6, however, the macroscopic data (see Fig. 6a) were not reliable for the transverse strains due to significant oscillations, which were caused by the high uncertainty in the macroscopic DIC system at such a low strain level. The microscopic data (see Fig. 6b), on the other hand, provided significantly more consistent and smooth curves at even lower strain levels. This high-resolution data set contributed to characterizing the in-plane Poisson's ratio of the material, which is one of the major inputs for the MAT213 model.

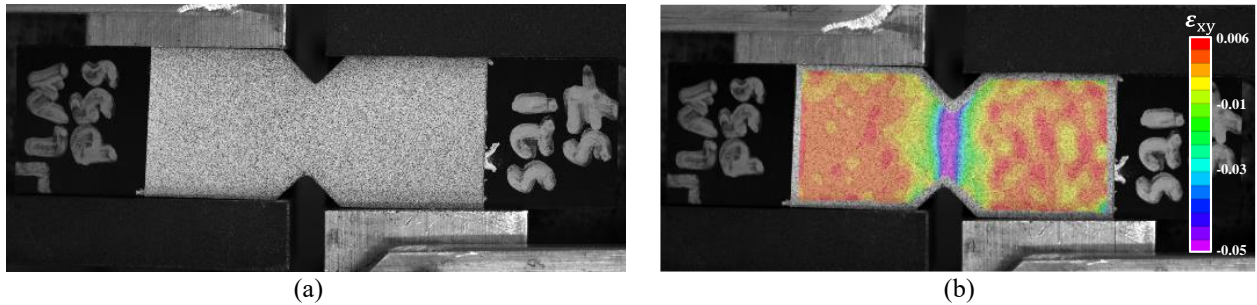

**Figure 4. V-notched beam test in 90° direction. (a) An image of a speckled V-notched coupon in the test fixture. (b) DIC analysis result of the image for shear strains.**

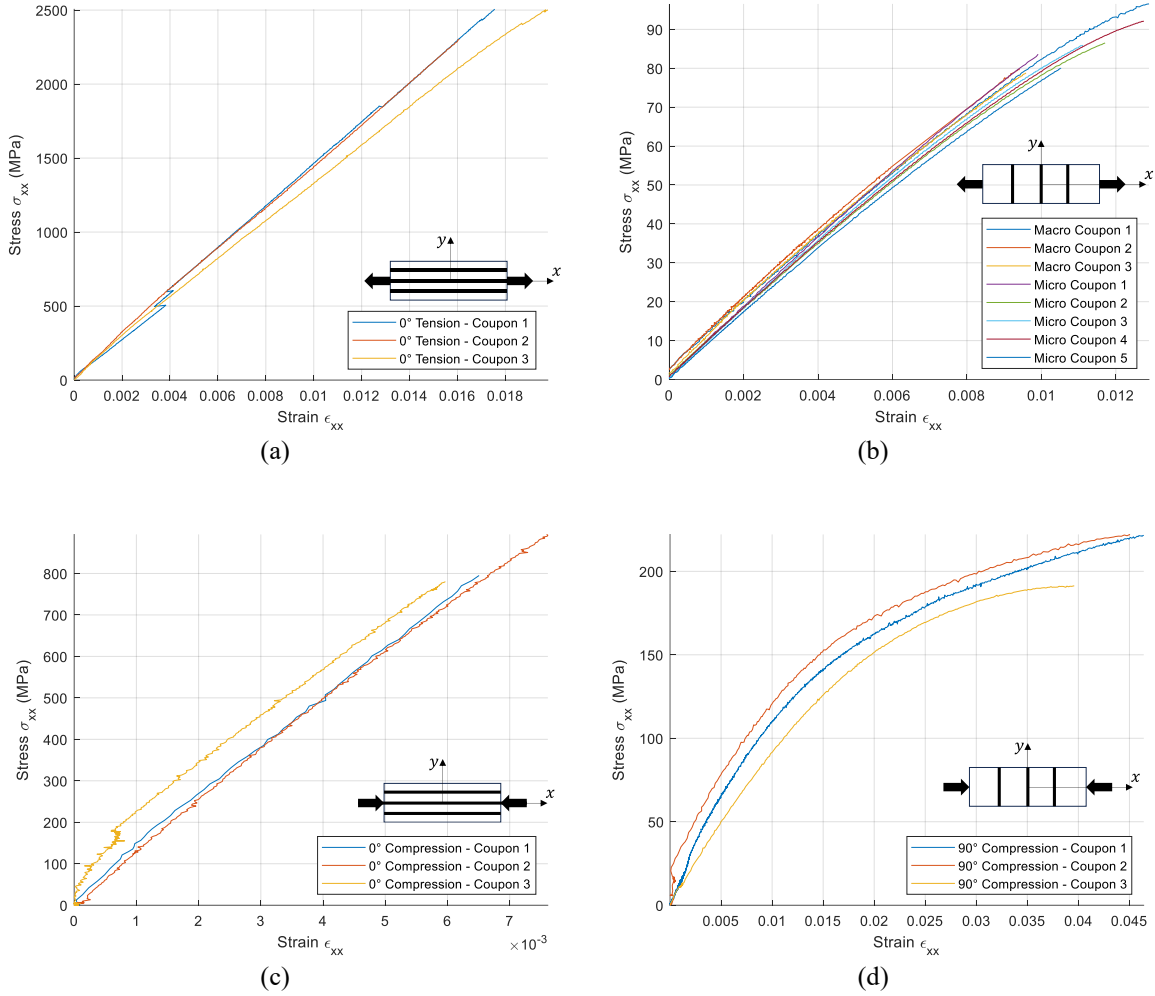

**Figure 5.** Full-scale stress-strain curves for longitudinal strains. (a) 0° unidirectional tension tests. (b) 90° unidirectional tension tests. (c) 0° unidirectional compression tests. (d) 90° unidirectional compression tests.

The material properties obtained from the stress-strain curves are tabulated with the NCAMP data in Table 2. The comparisons in the table were made based on the room-temperature test results on the individual test summary sheets in the NCAMP report; however, no data on the strengths of 0° unidirectional compression tests and the modulus values of notched in-plane shear tests were reported on the sheets. The report included the 0.2%- and 5%-offset strength values from tension tests of ±45° laminates; however, comparisons were not made in the table due to the inconsistency in offset strength measurements. The shear strength properties were instead characterized from the V-notched beam tests and were compared with the NCAMP data. It also needs to be noted that the in-plane shear data in the NCAMP report were not specified for fiber directions and thus were compared with the V-notched beam test data in both the 0° and 90° directions. As shown in the table, the test data overall showed good agreement with the NCAMP data.

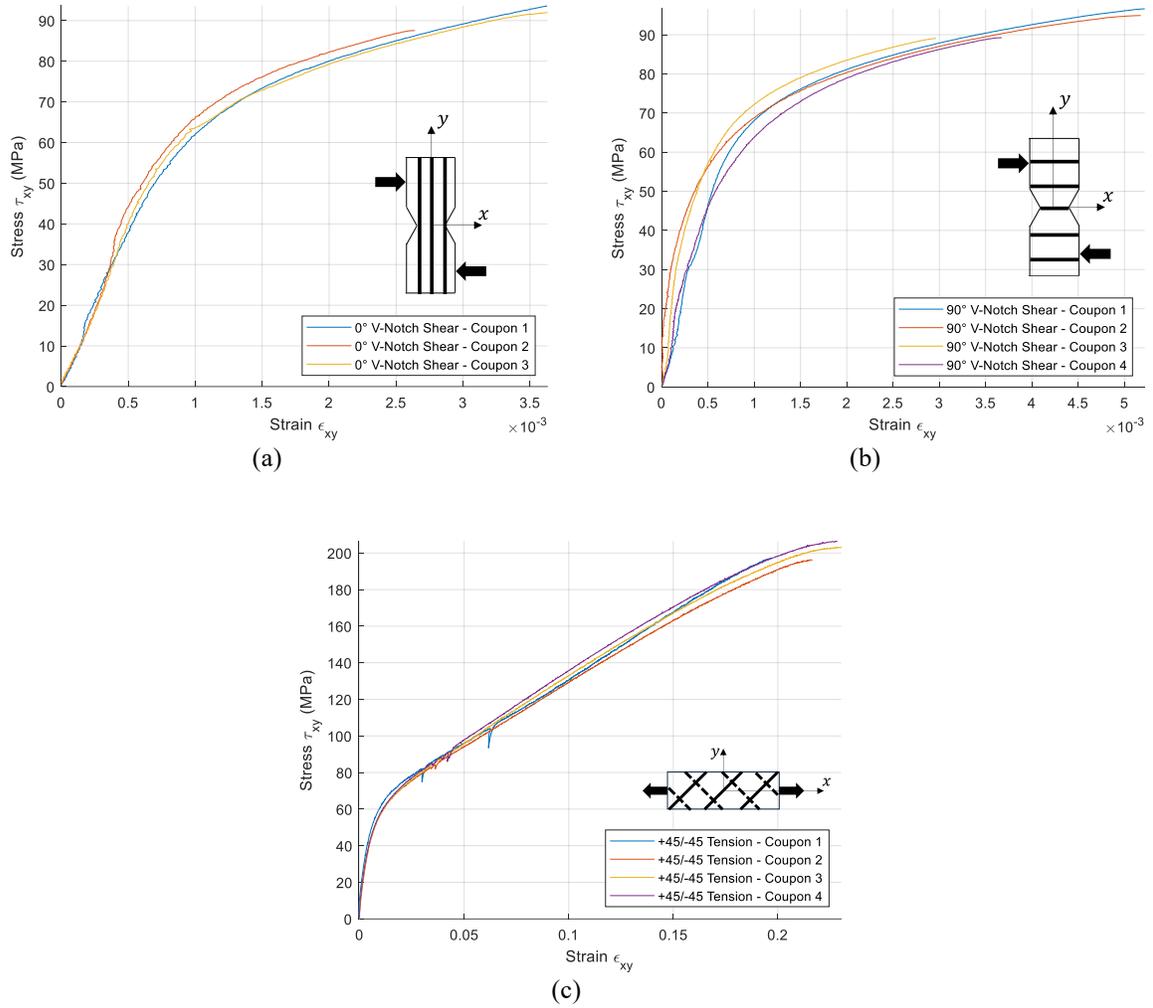

**Figure 6. Full-scale stress-strain curves for shear strains. (a) V-notched beam tests in 0° direction. (b) V-notched beam tests in 90° direction. (c) Tension tests of ±45° laminates.**

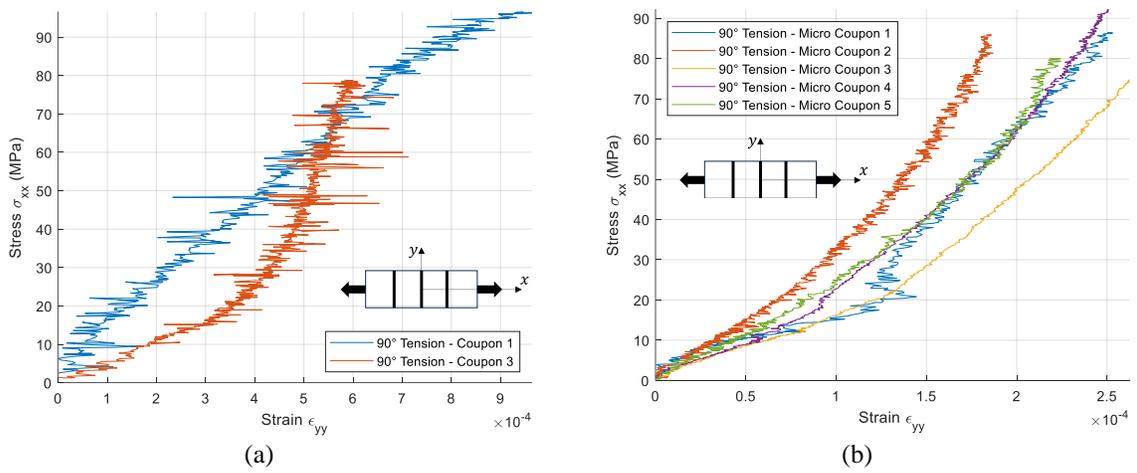

**Figure 7. Full-scale stress-strain curves for transverse strains under 90° unidirectional tension. (a) Macroscopic data. (b) Microscopic data.**

Table 2. Test summary

| Test types | | Modulus (GPa) | | | Strength (MPa) | | |
|---|---|---|---|---|---|---|---|
| | | Test | NCAMP* | % difference | Test | NCAMP* | % difference |
| 0° unidirectional tension tests | Mean | 130.3 | 130.5 | -0.19% | 2,437.1 | 2422.1 | -0.62% |
| | Min. | 125.5 | 126.9 | -1.13% | 2,303.4 | 2124.3 | 8.43% |
| | Max. | 135.8 | 138.0 | -1.62% | 2,506.0 | 2747.6 | -8.79% |
| 90° unidirectional tension tests (macroscopic) | Mean | 8.80 | 9.09 | -3.21% | 85.1 | 94.3 | -9.78% |
| | Min. | 8.56 | 8.75 | -2.17% | 78.8 | 84.5 | -6.78% |
| | Max. | 9.17 | 9.68 | -5.27% | 96.6 | 100.6 | -3.97% |
| 90° unidirectional tension tests (microscopic) | Mean | 8.60 | 9.09 | -5.39% | 85.7 | 94.3 | -9.12% |
| | Min. | 8.32 | 8.75 | -4.91% | 80.0 | 84.5 | -5.36% |
| | Max. | 8.85 | 9.68 | -8.58% | 92.2 | 100.6 | -8.34% |
| 0° unidirectional compression tests | Mean | 117.95 | 116.80 | 0.99% | 823.3 | - | - |
| | Min. | 117.48 | 112.59 | 4.34% | 780.5 | - | - |
| | Max. | 118.59 | 121.00 | -1.99% | 894.3 | - | - |
| 90° unidirectional compression tests | Mean | 9.67 | 9.24 | 4.63% | 211.7 | 212.1 | -0.20% |
| | Min. | 9.44 | 8.94 | 5.66% | 191.3 | 204.8 | -6.61% |
| | Max. | 9.81 | 9.53 | 3.00% | 222.1 | 219.9 | 1.01% |
| Tension tests of ±45° laminates | Mean | 4.38 | 4.65 | -5.80% | - | - | - |
| | Min. | 4.23 | 4.44 | -4.69% | - | - | - |
| | Max. | 4.59 | 4.83 | -5.02% | - | - | - |
| V-notched beam tests in 0° direction | Mean | - | - | - | 91.0 | 97.2 | -6.35% |
| | Min. | - | - | - | 87.6 | 90.7 | -3.50% |
| | Max. | - | - | - | 93.6 | 100.7 | -7.02% |
| V-notched beam tests in 90° direction | Mean | - | - | - | 92.5 | 97.2 | -4.81% |
| | Min. | - | - | - | 89.1 | 90.7 | -1.85% |
| | Max. | - | - | - | 96.9 | 100.7 | -3.71% |

*The values were taken from the individual test summary sheets in the NCAMP report [4].

## IV. Analytical Simulations

The modeling work is still in progress but some of the preliminary simulation results are presented in this section. The input parameters used for the MAT213 simulations are shown in Fig. 8. The material properties in Table 2 were used for the input module shown in Fig. 8a. For the full-scale stress-strain data input module (see Fig. 8b), a smooth and continuous data set is required in the form of a tabulated input. Thus, the curve-fitting process is an important step for MAT213 modeling. Following the work in Ref. [2], the flow rule coefficients were calculated using the experimental data and were calibrated to match the data.

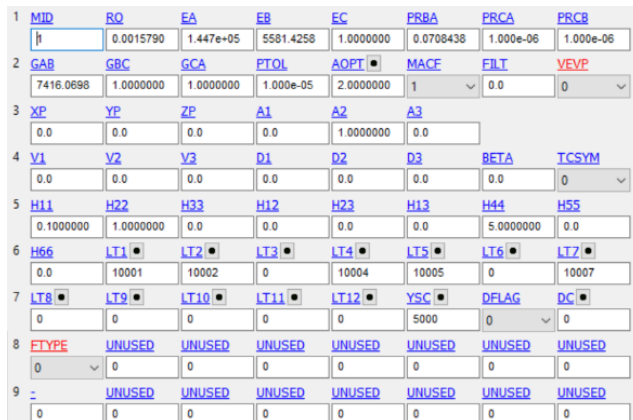
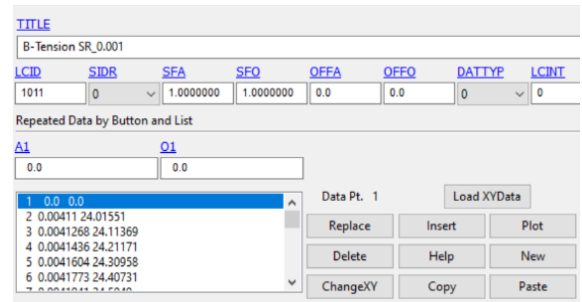

(a) (b)

**Figure 8.** Data input for the LS-DYNA MAT213 model. (a) Screenshot of the material property input module. (b) Screenshot of the full-scale stress-strain data input module.

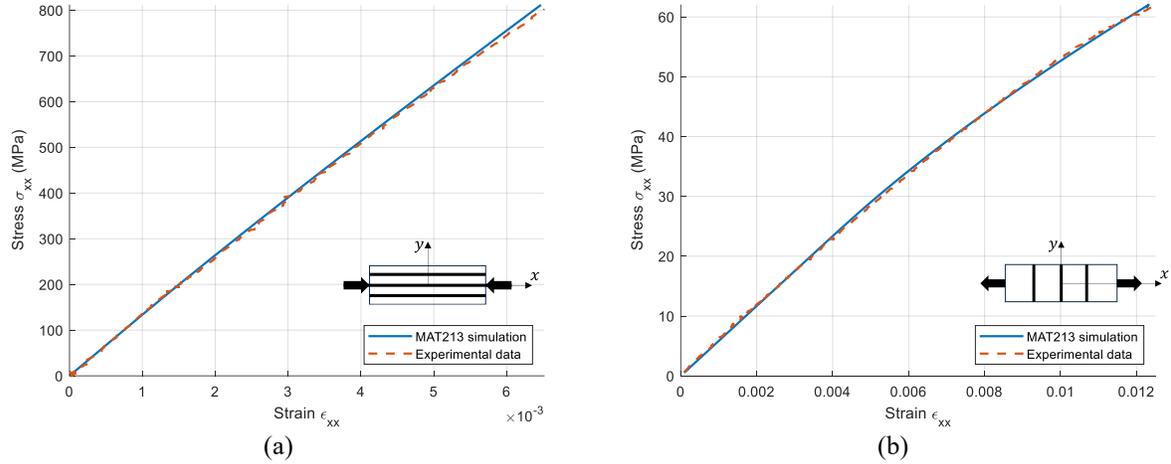

**Figure 9. Coupon-level simulation results obtained from the LS-DYNA MAT213 model. (a) 0° unidirectional compression test. (b) 90° unidirectional tension test.**

Single-element simulations were first performed in MAT002, a linear orthotropic material model available in LS-DYNA, before progressing to the considerably more advanced MAT213 model. Single-element modeling served to validate simple material properties such as in-plane normal and shear moduli, and Poisson's ratios. After validating the model for these properties, coupon-level simulations were conducted using the MAT213 model. Various LS-DYNA input parameters including experimental data inputs and calculated plasticity flow rule coefficients (see the H11, H22, and H44 values in Fig. 7a) were validated and confirmed in this step. Some of the simulation results for the experimental data shown in the previous section are presented in Fig. 8. It needs to be noted that these results were obtained using a deformation sub-model, and future work will be focused on developing a damage sub-model. More details on the modeling methodology and validation process will be presented with simulation results in a future paper.

## V. Conclusion

This paper was focused on developing and applying an experimental framework to obtain inputs for the LS-DYNA MAT213 model. The inputs include material properties, tabulated stress-strain curves, and non-associative plasticity flow rule coefficients. A specific composite material, T700G/LM-PAEK, was chosen for this study. For the in-plane tests of this material, the multi-scale DIC systems were applied. The microscopic DIC system provided high-resolution data sets and was particularly useful to obtain reliable and consistent data in small-strain regimes. Full-scale stress-strain curves were obtained, and the experimental data overall showed good agreement with the properties reported by NCAMP [4]. Some of the initial modeling efforts were also presented with simulation results obtained using a deformation sub-model in the MAT213 model. In future work, more efforts will be made to develop a damage sub-model for the material. For the characterization of post-peak stress-degradation, mode-II interlaminar fracture in the material will experimentally be characterized. Damage initiation and propagation in geometrically scaled specimens will be captured using the multi-scale DIC systems. A damage sub-model will be built and calibrated based on this experimental data set along with available high-strain-rate impact test data. This work will contribute to developing a high-fidelity simulation tool based on the MAT213 model for composite structures subjected to rate-dependent, multi-axial impact damages.

## Acknowledgments

The authors gratefully acknowledge the support of NASA (Award Number: MS-80NSSC22M0206), Sandi Miller at the NASA Glenn Research Center, and Joe Capriotti at the Advanced Composites Institute, Mississippi State University.